\documentclass[iop]{emulateapj}

\usepackage{graphicx}
\usepackage{apjfonts}
\usepackage{natbib}
\slugcomment{Submitted to ApJ}
\shorttitle{Assessing the behavior of modern solar magnetographs}
\shortauthors{Del Toro Iniesta \& Mart\'{\i}nez Pillet}

\newcommand{\degree}{\ensuremath{^\circ}\/}
\newcommand{\vect}[1]{\mathbf{#1}}
\usepackage{color}

\def\fdg{\hbox{$.\!\!^\circ$}}

\def\bbbone{{1\mskip-7mu 1}}

\begin{document}

\title{Assessing the behavior of modern solar magnetographs and spectropolarimeters}

\author{J.C.\ del Toro Iniesta}
\affil{Instituto de Astrof\'{\i}sica de Andaluc\'{\i}a (CSIC), Apdo.\ de Correos 3004, E-18080 Granada, Spain, jti@iaa.es}
\and
\author{V. Mart\'{\i}nez Pillet}
\affil{Instituto de Astrof\'{\i}sica de Canarias, V\'{\i}a L\'actea, s/n, E-28200 La Laguna, Spain, vmp@iac.es}

\begin{abstract}
The design and later use of modern spectropolarimeters and magnetographs
require a number of tolerance specifications that allow the developers to build
the instrument and then the scientists to interpret the data accuracy. Such
specifications depend both on device-specific features and on the physical
assumptions underlying the particular measurement technique. Here we discuss
general properties of every magnetograph, as the detectability thresholds for
the vector magnetic field and the line-of-sight velocity, as well as specific
properties of a given type of instrument, namely that based on a pair of
nematic liquid crystal variable retarders and a Fabry-P\'erot etalon (or
several) for carrying out the light polarization modulation and spectral
analysis, respectively. We derive formulae that give the detection thresholds
in terms of the signal-to-noise ratio of the observations and the polarimetric
efficiencies of the instrument. Relationships are also established between
inaccuracies in the solar physical quantities and instabilities in the
instrument parameters. Such relationships allow, for example, to translate
scientific requirements for the velocity or the magnetic field into
requirements for temperature or voltage stability. We also demonstrate that
this type of magnetograph can theoretically reach the optimum polarimetric
efficiencies of an ideal polarimeter, regardless of the optics in between the
modulator and the analyzer. Such optics induces changes in the instrument
parameters that are calculated too.  \end{abstract}


\section{Introduction}
\label{sec:intro}

Spectropolarimetry and magnetography have become two of the most useful tools in solar
physics because they provide the deepest analysis one can make of light. Solar
information is encoded in the spectrum of the Stokes parameters. We measure
this spectrum and infer solar quantities from it. Recently, less and less
conceptual differences exist between spectropolarimeters and magnetographs
except for the specific devices: the formers usually include those instruments
using a scanning spectrograph; the latter usually employ a bidimensional
filtergraph like a Fabry-P\'erot etalon. Some decades ago magnetographs were
only able to sample one or two wavelengths across a spectral line; nowadays,
new technologies provide a better wavelength sampling, thus enabling the
scientists to interpret the data in terms of sophisticated inversion techniques
of the radiative transfer equation, a procedure similar to the one regularly
used with spectropolarimeters. Some of the instruments mentioned below enter
this category. Modern solar spectropolarimeters and magnetographs are often
vectorial because all four Stokes parameters of the light spectrum are
measured. Longitudinal magnetography (i.e., Stokes $I \pm V$) can be
interesting for some specific applications, but the partial analysis is usually
included (if possible) as a particular case of the more general, full-Stokes
polarimetry. Some of these modern instruments have been recently built
or are currently in operation (e.g., the Tenerife Infrared Polarimeter, TIP,
\citeauthor{martinez+etal1999} \citeyear{martinez+etal1999},
\citeauthor{collados+etal2007} \citeyear{collados+etal2007}; the
Diffraction-Limited Spectro-Polarimeter, DLSP,
\citeauthor{sankarasubramanian+etal2003}
\citeyear{sankarasubramanian+etal2003}; the air-spaced Fabry-Perot based CRISP instrument, \citeauthor{2008ApJ...689L..69S} 
\citeyear{2008ApJ...689L..69S}; the spectropolarimeter, SP,
\citeauthor{2001ASPC..236...33L} \citeyear{2001ASPC..236...33L}, for the  {\em
Hinode} mission, \citeauthor{kosugi+etal2007} \citeyear{kosugi+etal2007};
CRISP, \citeauthor{2008...ESPFreib...N} \citeyear{2008...ESPFreib...N}; the
Visible Imaging Polarimeter, VIP, \citeauthor{2010A&A...520A.115B}
\citeyear{2010A&A...520A.115B}; the Imaging Magnetograph eXperiment, IMaX,
\citeauthor {2011SoPh...268...57M} \citeyear{2011SoPh...268...57M}, for the
{\em Sunrise} mission, \citeauthor{2011SoPh..268....1B}
\citeyear{2011SoPh..268....1B}; and the Helioseismic and Magnetic Imager, HMI,
\citeauthor{2003ASPC..307..131G} \citeyear{2003ASPC..307..131G}, for the {\em
Solar Dynamics Observatory} mission, \citeauthor{2000SPD....31.0704T}
\citeyear{2000SPD....31.0704T}). Some other are being designed and built for
near future operation and missions (e.g., the Polarimetric and Helioseismic
Imager, SO/PHI, [formerly called VIM, \citeauthor{2007ESApub.M}
\citeyear{2007ESApub.M}] for the {\em Solar Orbiter} mission,
\citeauthor{2005AdSpR..36.1360M} \citeyear{2005AdSpR..36.1360M}). 

The common interest of users of these instruments is centered in vector
magnetic fields (of components $B$, $\gamma$, and $\phi$) and line-of-sight
(LOS) velocities ($v_{\rm LOS}$). Some spectropolarimeters can provide
information on temperatures as well (and eventually on another thermodynamical
quantity) but that feature is not common to all of them. Therefore, assessing
the magnetograph capabilities in terms of their accuracy for retrieving
magnetographic and tachographic quantities is in order since such an analysis
can diagnose how far reaching is our current knowledge of the solar dynamics
and magnetism. The diagnostics is relevant both for the design of new
instruments in order to maximize their performances and for the analysis of
uncertainties in data coming from currently operating devices. General
considerations can obviously not be made but a few. Specifically, we here study
in Sect.\ \ref{sec:threshold} the detection thresholds induced by random noise
on the inferred longitudinal and transverse components of the magnetic field;
in the particular case of photon-induced noise we also find uncertainty
formulas. Both thresholds and relative uncertainties are obtained in terms of
the signal-to-noise ratio of the observations and of the polarimetric
efficiencies of the instrument. Since such efficiencies vary from instrument to
instrument, at that point, the analysis concentrates in a particular type of
magnetograph, namely that consisting of two nematic liquid crystal variable
retarders (LCVRs) as the polarization modulator and a Fabry-P\'erot etalon. In
Sect.\ \ref{sec:optimpol}, we demonstrate that these polarimeters can reach the
theoretically optimum efficiencies no matter the optics behind the modulator,
including the etalon. The way for calculating the required retardances for the
two LCVRs are explained along with a number of rules and periodicities in the
solutions. Section \ref{sec:instabilities} analyzes these instruments in terms
of the influence of temperature and voltage instabilities, as well as of
thickness inhomogeneities (roughness), of both the LCVRs and the etalon(s), on
the final magnetographic and tachographic measurements. Finally, Sect.\
\ref{sec:conclu} summarizes the results.

\section{The thresholding action of random noise}
\label{sec:threshold}

Most astrophysical measurements are nothing but photon counting. Their
accuracy, therefore, depends on photometric accuracy, that is, on a battle
between our ability to detect changes in the solar (stellar) physical
quantities and the noise that hide such changes. The key concept is {\em
changes}: we need to discern if a given quantity like the magnetic field
strength, $B$, or the line-of-sight (LOS) velocity, $v_{\rm LOS}$, varies among
pixels: whether or not it is greater or smaller than in the neighbor zones. The
only tool we have to gauge these changes is the observable Stokes
parameter changes that are linked to them through the response functions.
Discussing response functions is out of the scope of this paper as they have
been extensively analyzed elsewhere
\citep[e.g.,][]{2010ApJ...711..312D,2007A&A...462.1137O,2003isp..book.....D,1996SoPh..164..169D,
1994A&A...283..129R,1977A&A....56..111L,1971SoPh...20....3M}.
As explained by \citet{2010arXiv1010.0504D}, purely phenomenological approaches
\citep[e.g.,][]{2005A&A...439..687C} are also valid to establish a relationship
between changes in the physical quantities and the Stokes parameters. Here we
shall concentrate on noise; on how it establishes the minimum threshold below
which no signal changes can be detected.  In spectropolarimetry, the customary
estimate for noise (and thus for the detection threshold) is the standard
deviation of the continuum signal because polarization is assumed to be
constant at continuum wavelengths. Therefore, the noise is calculated either
over a continuum window in a given spatial pixel or over all the spatial pixels
of a map in a given continuum wavelength sample. Both estimates should agree as
it is the case in most observations. If we call $\vect{S} \equiv (S_1, S_2,
S_3, S_4)$ the (pseudo-)vector of Stokes parameters and denote by $\sigma_i$,
with $i=1,2,3,4$, the standard deviation of each Stokes parameter, then the
signal-to-noise ratio in each parameter, $(S/N)_i$, is defined as the inverse
of this deviation in units of the continuum intensity:
\begin{equation}
\label{eq:snratio}
(S/N)_i = \left( \frac{S_1}{\sigma_i} \right)_{\!\!\!\rm c},
\end{equation}
where index c refers to continuum. Thus, when we say that our observations
have, for example, a $S/N = 1000$, we mean that when signals in a given (non
specified) Stokes parameter are greater than $10^{-3} S_{1,{\rm c}}$  can be
detected and this is certainly valid for that parameter but {\em not
necessarily} for the others. As a matter of fact, if noise is random (or
uncorrelated with the signal) and can be represented by a Gaussian distribution
\citep{2009ASPC..405..371K}, according to \cite{2000ApOpt..39.1637D}, 
\begin{equation}
\label{eq:noisequv}
\sigma_i = \frac{\varepsilon_1}{\varepsilon_i} \sigma_1, \,\,\, i=1,2,3,4,
\end{equation}
where $\varepsilon_i$ stands for each one of the so-called polarimetric
efficiencies of the instrument. The polarimetric efficiencies depend
in a non-linear way on the modulation matrix elements (cf. Eq. \ref{eq:defeff})
that on their turn come from the first row of the polarimeter Mueller matrix
\citep{2003isp..book.....D}. Since all the efficiencies are necessarily
less than the first (that of the intensity), Eq.\ (\ref{eq:noisequv}) means
that the noise is always larger in the polarization parameters than in the
intensity. Then one can easily see that \citep[see][]{martinez+etal1999}
\begin{equation}
\label{eq:stonquv}
(S/N)_i = \frac{\varepsilon_i}{\varepsilon_1} (S/N)_1, \,\,\, i=2,3,4,
\end{equation}
that is, that the signal-to-noise ratio for Stokes $S_2$, $S_3$, and $S_4$ is
always less than that for Stokes $S_1$. Let us point out here, however, that
other systematic (or instrumental) errors like those introduced by flat
fielding of images may invalidate the above equation. We are explicitly
discarding these other sources of noise from our analysis.

The Stokes parameters cannot be measured with single exposures. Instead, for
vector polarimetry, a number $N_{p} \geq 4$ of single detector shots are
recorded each providing a linear combination of all the four Stokes parameters.
The set of $N_{p}$ individual measurements constitutes a modulation cycle that
is characteristic of an  instrument mode of operation. After demodulation, that
is, after solving the set of $N_{p}$ linear equations made up of the individual
exposures, the Stokes vector is measured. To increase the signal-to-noise ratio
of the measurement, many instruments use $N_{a}$ accumulations, that is, repeat
the modulation cycle $N_{a}$ times and the corresponding polarization images
are added together. Since the degree of polarization of the incoming typical
solar beam is fairly small, each single shot usually has the same light levels
and, hence, the same (photon-noise-dominated) signal-to-noise ratio $s/n$.
Thus, the signal-to-noise ratio of $S_1$ is related to the single-shot $s/n$
\citep[see][for an illustrative description]{2007ESApub.M} through 
\begin{equation}
\label{eq:individualglobal}
(S/N)_1= (s/n) \, {\varepsilon_1} \sqrt{N_p N_a},
\end{equation} 
because $S_{1}$ is often retrieved from the sum of all the accumulated images
for all the polarization exposures of the given modulation scheme. An advisable
practice for characterizing the signal-to-noise ratio of an instrument is to
always refer to that of intensity and equate $S/N = (S/N)_1$. This is
convenient because there is only one intensity while the other polarization
Stokes parameters are three and one would need to specify which one is meant
each time. Let us remark, however, that this convention is not universal and
some authors, always interested in the polarization features, think of and
speak about some of the other three Stokes parameter signal-to-noise ratios.
Such an alternative convention makes sense if one takes into account the
differential character of polarization measurements: demodulation implies that
Stokes $S_{2,3,4}$ are essentially retrieved from image differences; hence, any
systematic error like that produced by flat fielding is naturally mitigated (or
eventually cancels out). On the contrary, the additive character of Stokes
$S_{1}$  implies that intensity noise can be higher than simple photon noise.
We shall hereafter follow the $S/N = (S/N)_1$ convention in the paper for the
sake of simplicity in the description and in the equations (thus, explicitly
neglecting systematic errors). When people is more interested in the other
three Stokes parameter signal-to-noise ratios, Eq.\ (\ref{eq:stonquv}) provides
the obvious help. 

As demonstrated by  \cite{2000ApOpt..39.1637D}, the maximum efficiencies that
an ideal system can have are ${\vect{\varepsilon}} = (1, 1/\sqrt{3},
1/\sqrt{3}, 1/\sqrt{3})$, if all the three last ones are equal. Therefore, the
relationship $(S/N)_i \leq  1/\sqrt{3}\,  (S/N)$, $i=2,3,4$, holds for any
random-noise-dominated polarimetric system.\footnote{There may be polarimeters
that are designed to measure not all four Stokes parameters or that aim at
better accuracies for given Stokes parameters. In such cases some (never all)
efficiencies can be greater than $1/\sqrt{3}$. In what follows, however, we
assume that our interest is the same for $S_2$, $S_3$, and $S_4$.} Or, in terms
of Eq.\ (\ref{eq:noisequv}), necessarily,
\begin{equation}
\label{eq:sigmai}
\sigma_i \geq \sqrt{3}\, \sigma_1.
\end{equation}
Equation (\ref{eq:sigmai}) means that the detection threshold is bigger for the
polarization parameters than for the intensity. {\em Detectability is smaller
in polarimetry than in pure photometry}.

An important parameter describing the state of any beam of light is its degree of polarization
\begin{equation}
\label{eq:poldegree}
p^2 \equiv \frac{1}{S_1^2} \sum_{i=2}^4 S_i^2.
\end{equation}
Now the question naturally arises as to what is the minimum detectable degree
of polarization by a given polarimetric system. If the uncertainties in the
Stokes parameters are uncorrelated (and this should be especially true when
considering a statistics on all the pixels of an image), error propagation in
Eq.\ (\ref{eq:poldegree}) gives
\begin{equation}
\label{eq:errdegree1}
\sigma_{p^2}^2 = \sum_{i=1}^4 \left( \frac{\partial p^2}{\partial s_i} \right)^2 \sigma_{s_i}^2,
\end{equation}
where, for convenience we have made $s_i \equiv S_i^2$, $i=1,2,3,4$. Now, since $\sigma_{s_i} = 2 S_i \sigma_i$, it is easy to see that 
\begin{equation}
\label{eq:errdegree2}
\sigma_{p^2}^2 = \frac{4}{S_1^4} \left( p^4 S_1^2 \sigma_1^2 + \sum_{i=2}^4 S_i^2 \sigma_{i}^2 \right)
\end{equation}
and, according to Eq.\ (\ref{eq:noisequv}) and since $\sigma_{p^2} = 2p\, \sigma_p$, one can write 
\begin{equation}
\label{eq:errdegree3}
\frac{\sigma_p^2}{p^2} =  \left( 1 +  \frac{1}{p^4} \sum_{i=2}^4 \frac{S_i^2}{S_1^2} 
\frac{\varepsilon_1^2}{\varepsilon_i^2} \right) \frac{\sigma_1^2}{S_1^2}.
\end{equation}
Now, if all the three last efficiencies are the same and certainly less than their maxima, we finally obtain
\begin{equation}
\label{stondegree}
\frac{\sigma_p}{p} \geq \frac{\sqrt{1 + \frac{3}{p^2}}}{S/N},
\end{equation}
an inequality already published by \citet{2010AN....331..558D},
\citet{2011SoPh...268...57M}, and \citet{2010arXiv1010.0504D} without
demonstration.

Our instruments are aimed at measuring magnetic fields and velocities.
Therefore, any reasonable design should include lower limits for these
quantities within the overall error budget. Detectability thresholds for the
Stokes parameters imply thresholds for the magnetograph and tachograph signals
as well. The rest of this section is devoted to estimate them. Of course, any
estimation that one can make depends not only on the instrument but also on the
inference technique. Most modern magnetographs use inversion of the radiative
transfer equation to infer values for both the magnetic field vector and the
plasma velocity. These inferences involve all four Stokes parameters and,
hence, should be more accurate than those using just one or two of them.
However, for the sake of clarity in the analytical derivation, we shall
consider errors induced in  the magnetographic and tachographic formulas
(\ref{eq:blon}), (\ref{eq:btran}), and (\ref{eq:tachometer}). 

Using classical magnetographic formulas, the longitudinal and transverse components 
of the magnetic field are given by
\begin{equation}
\label{eq:blon}
B_{\rm lon} = k_{\rm lon} \frac{V_s}{S_{1,\rm c}}
\end{equation}
and 
\begin{equation}
\label{eq:btran}
B_{\rm tran} = k_{\rm tran} \sqrt{\frac{L_s}{S_{1,\rm c}}},
\end{equation}
where $k_{\rm lon}$ and $k_{\rm tran}$ are (model-dependent) calibration
coefficients and $V_s$ and $L_s$ are the circular and linear polarization
signals calculated as
\begin{equation}
\label{eq:circpol}
V_s \equiv \frac{1}{n_\lambda} \sum_{i=1}^{n_\lambda} a_i \, S_{4,i},
\end{equation}
where $a_i = 1$ or $-1$ depending on whether the sample is to the blue
(including the zero shift) or the red side of the central wavelength of the
line, and 
\begin{equation}
\label{eq:linpol}
L_s \equiv \frac{1}{n_\lambda} \sum_{i=1}^{n_\lambda} \sqrt{ S_{2,i}^2 + S_{3,i}^2 }.
\end{equation}
In the above equations, $n_\lambda$ stands for the number of wavelength
samples. If we now assume that the minimum polarization signals are $V_s =
\sigma_4$ and $L_s = \sigma_2 = \sigma_3$, the minimum detectable thresholds
are 
\begin{equation}
\label{eq:uncerblon}
\delta (B_{\rm lon}) = \frac{ k_{\rm lon}}{S/N}\, \frac{\varepsilon_{1}}{\varepsilon_{4}}
\end{equation}
and
\begin{equation}
\label{eq:uncertran}
\delta (B_{\rm tran}) =  k_{\rm tran} \sqrt{\frac{\varepsilon_{1}/\varepsilon_{2}}{S/N}} = k_{\rm tran}
\sqrt{\frac{\varepsilon_{1}/\varepsilon_{3}}{S/N}}.
\end{equation}
Expressions (\ref{eq:uncerblon}) and (\ref{eq:uncertran}) give the explicit
dependence of magnetic detectability thresholds in terms of the instrument
efficiencies and are very useful in practice. For example, if we use the
calibration constants for IMaX quoted in \citet{2011SoPh...268...57M}, assuming
that the maximum polarimetric efficiencies have been reached, and typical
signal-to-noise ratios of 1700 ($\sim1000$ for $S_2, S_3$, and $S_4$), the
minimum longitudinal and transverse components of the magnetic field detectable
with that magnetograph are 5 and 80 G, respectively. 

As far as the velocity is concerned, we shall assume that the Fourier
tachometer technique \citep{1978...JOSO...B,1981...NSO...B,1992PhDT.........3F}
is used:
\begin{equation}
\label{eq:tachometer}
v_{\rm LOS} = \frac{2{\rm c}\, \delta\lambda}{\pi\lambda_0} \, 
\arctan \frac{S_{1,-9}+S_{1,-3}-S_{1,+3}-S_{1,+9}}{S_{1,-9}-S_{1,-3}-S_{1,+3}+S_{1,+9}},
\end{equation}
where c is the speed of light, $\delta\lambda$ is the spectral resolution of
the instrument, and $\lambda_0$ is the central wavelength of the line; $-9, -3,
+3, +9$ stand for the sample wavelengths of the intensity, measured in
picometers with respect to $\lambda_{0}$. Let us assume that the minimum
detectable difference between symmetric wavelength samples 
(such as $S_{1,-3}-S_{1,+3}$) due to LOS velocity
shifts is $\sigma_1$. Then, if the difference between the samples at the same
flank of the line is approximated by $~1/2 \,\, S_{1,\rm c}$, the minimum 
detectable LOS velocity can be approximated by
\begin{equation}
\label{eq:tachothreshold}
\delta(v_{\rm LOS}) \simeq \frac{2{\rm c}\, \delta\lambda}{\pi\lambda_0} \, \arctan \frac{2}{S/N}.
\end{equation}
Likewise Eqs. (\ref{stondegree}), (\ref{eq:uncerblon}) and
(\ref{eq:uncertran}), this new expression (\ref{eq:tachothreshold}) relates the
velocity threshold with the signal-to-noise ratio of the instrument. If we use
again IMaX values ($\delta\lambda = 8.5$ pm; $\lambda_0 = 525.02$ nm) and
assume $S/N = 1700$, the minimum detectable LOS velocity change is roughly 4
m$\,$s$^{-1}$.

\subsection{Uncertainties induced by photon noise}
\label{sec:uncer}

Fluctuations in the light levels due to photon statistics necessarily imply
variances in the Stokes parameters that in the end induce uncertainties in the
measured physical quantities, $B_{\rm lon}$, $B_{\rm tran}$, and $v_{\rm LOS}$. In this
section, we are going to establish a relationship between those variances and
uncertainties. Note that we discard for the moment any random fluctuation in
the instrument that will be dealt with in Sect.\ \ref{sec:instabilities}. 

Error propagation in Eq.\ (\ref{eq:blon}) easily yields
\begin{equation}
\label{eq:errblos}
\frac{\sigma^2_{B_{\rm lon}}}{B_{\rm lon}^2} = \frac{\sigma_4^2}{n_{\lambda} V_s^2} + \left( \frac{1}{S/N} \right)^2,
\end{equation}
because $\sigma^2_{V_s} = \sigma^2_4 / n_{\lambda}$. Using now Eqs.\ (\ref{eq:noisequv}), (\ref{eq:blon}), and (\ref{eq:snratio}), one obtains
\begin{equation}
\label{eq:errblosdef}
\frac{\sigma^2_{B_{\rm lon}}}{B_{\rm lon}^2} = \left( \frac{k_{\rm lon}^2}{n_{\lambda} B_{\rm lon}^2} \frac{\varepsilon_{1}^2}{\varepsilon_{4}^2} +1 \right) \frac{1}{(S/N)^2},
\end{equation}
that relates the $B_{\rm lon}$ relative error with itself, the signal-to-noise
ratio of the observations, and the polarimetric efficiencies. On its turn,
error propagation in Eq.\ (\ref{eq:btran}) gives
\begin{equation}
\label{eq:errbtran}
\frac{\sigma^2_{B_{\rm tran}}}{B_{\rm tran}^2} = \frac{1}{4} \left[ \frac{\sigma_{L_{s}}^2}{L_{s}^2} + \left( \frac{1}{S/N} \right)^2 \right].
\end{equation}
Now, since the variances for Stokes $S_{2}$ and Stokes $S_{3}$ should be
approximately the same, $\sigma_{L_{s}}^2 \simeq (1/2n_{\lambda}) (\sigma_{2}^2
+ \sigma_{3}^2)$ and, using Eqs.\ (\ref{eq:noisequv}), (\ref{eq:btran}), and
(\ref{eq:snratio}), Eq.\ (\ref{eq:errbtran}) turns out to be
\begin{equation}
\label{eq:errbtrandef}
\frac{\sigma^2_{B_{\rm tran}}}{B_{\rm tran}^2} = \left[ \frac{k_{\rm tran}^4}{8 n_{\lambda} B_{\rm tran}^4} \left( \frac{\varepsilon_{1}^2}{\varepsilon_{2}^2} + \frac{\varepsilon_{1}^2}{\varepsilon_{3}^2} \right) + \frac{1}{4} \right] \frac{1}{(S/N)^2},
\end{equation}
that again relates the relevant magnetographic quantity relative error with
itself, the photon-induced signal-to-noise ratio of the observations, and the
polarimetric efficiencies of the instrument.

If we use the values for the calibration constants quoted in
\citet{2011SoPh...268...57M} for IMaX, $n_{\lambda} = 5$ for this instrument,
and assume that the maximum polarimetric efficiencies are reached, then the
estimated relative errors for $B_{\rm lon}$ and $B_{\rm tran}$ induced by a
photon noise of $S/N = 1700$ are of 2 and 15\%, respectively, for magnitudes in
either quantities of 100 G; for magnitudes of 1000 G, the relative errors drop
to 0.2 and 0.1\%, respectively. 

After a similar calculation for photon-noise-induced uncertainties in the tachographic formula (\ref{eq:tachometer}), one gets
\begin{equation}
\label{eq:noiseinducederrvel}
\sigma_{v_{\rm LOS}}^2 = \frac{4c^2(\delta\lambda)^2}{\pi^2 \lambda_0^2} \frac{2\sigma_1^2}{\Delta},
\end{equation}
where $\Delta = (S_{1,-9} - S_{1,+3})^2 + (S_{1,+9} - S_{1,-3})^2$, and the
variances of the Stokes $S_{1}$ samples are all assumed to be $\sigma_{1}$.
Note that the slight asymmetry between Eq.\ (\ref{eq:noiseinducederrvel}) and
Eqs.\ (\ref{eq:errblosdef}) and (\ref{eq:errbtrandef}) 
is not such as the ratio $\sigma_{1}^2/\Delta$ is a kind of inverse, square signal-to-noise ratio. A
numerical estimate for the IMaX instrument, and using the FTS spectrum by
\citet{1987ftp...book...B} to evaluate $\Delta$ for its Fe~{\sc i} line at
525.02 nm, we conclude that the photon-noise-induced uncertainty is 4
m$\,$s$^{-1}$.

\section{An optimum vector plus longitudinal polarimeter}
\label{sec:optimpol}

As explained by \citet{2004SPIE.5487.1152M}, a versatile polarimeter is
obtained through the combination of two nematic liquid crystal variable
retarders (LCVRs) with their optical axes properly oriented at 0\degree and
45\degree with the Stokes $S_2$ positive ($X$) direction. This is so because it
can provide optimum modulation schemes for both the vectorial and the
longitudinal $(S_1\pm S_4)$ polarization analyses by simply tuning the voltages
that change their retardances. The theoretical maximum efficiencies mentioned
above can in principle be reached by such an ideal polarimeter. We have assumed
these maximum efficiencies for our instruments so far. However, instrumental
effects may corrupt the measurement so that the final efficiencies are lower.
Let us see in this section what happens if some typical optical elements are
included between the modulator and the analyzer in the analysis.

The corrupting effect of the optical elements of an instrument in the final
polarization analysis is called instrumental polarization. It is well known
that those optical components acting on light after the polarization modulation
do not produce any instrumental polarization. However, nobody has yet
demonstrated whether optimum polarimetric efficiencies can still be reached no
matter the optics in between the modulator and the analyzer. In this section we
are going to show that this is the case with these two-LCVR-based polarimeters
because retardances can be fine tuned by simply changing the acting voltages.
This property certainly makes this type of polarimeters very versatile and
optimum for solar investigations. To understand the result, let us start by
demonstrating that, indeed, a polarimeter made up of two nematic LCVRs oriented
as above plus a linear analyzer can reach the optimum polarimetric
efficiencies.

According to \citet{2003isp..book.....D}, the modulation matrix of any
polarimetric system consists of rows that equal the first row of the system
Mueller matrix for each of the measurements. If ${\bf R} (\theta, \delta)$
stands for the Mueller matrix of a general retarder whose fast axis is at an
angle $\theta$ with the $X$ axis and whose retardance is $\delta$, our LCVR
Mueller matrices can be described by ${\bf M}_1 = {\bf R} (0,\rho_i)$ and ${\bf
M}_2 = {\bf R} (\pi/4,\tau_i)$, where $i=1,2,3,4$ is an index for each of the
four measurements. Hence, in our case, where the analyzer (of Mueller matrix
${\bf M}_4$) is a linear polarizer at 0\degree,\footnote{Dual-beam polarimeters
use a polarizing beam splitter as a double analyzer. Hence, another analyzer at
90\degree is indeed present simultaneously although the double calculation is
not necessary.} such a modulation matrix, disregarding a 1/2 gain factor, is
given by  \citep[see][]{2004SPIE.5487.1152M}
\begin{equation}
\label{eq:modulmat}
{\bf O} = \left( \begin{array}{llll}
1 & \cos \tau_1 & \sin \rho_1 \sin \tau_1 & -\cos\rho_1 \sin\tau_1 \\
1 & \cos \tau_2 & \sin \rho_2 \sin \tau_2 & -\cos\rho_2 \sin\tau_2 \\
1 & \cos \tau_3 & \sin \rho_3 \sin \tau_3 & -\cos\rho_3 \sin\tau_3 \\
1 & \cos \tau_4 & \sin \rho_4 \sin \tau_4 & -\cos\rho_4 \sin\tau_4 
\end{array} \right).
\end{equation}
As explained in  \citet{2000ApOpt..39.1637D}, if all the three last column
elements of ${\bf O}$ have a magnitude of $1/\sqrt{3}$ (with their signs
properly altered), then the modulation is optimum and the maximum efficiencies
are reached. $|\cos\tau| = 1/\sqrt{3}$ has the four independent solutions $\tau
= 54\fdg 736, 125\fdg 264, 234\fdg 736,$ and $305\fdg 264$. With them, $|\sin
\rho \,  \sin \tau| = 1/\sqrt{3}$ is equivalent to $|\sin \rho| = \sqrt{2}/2$
that has four independent solutions as well: $\rho = 45\degree, 135\degree,
225\degree$, and $315\degree$. The verification of the above two equations
ensures the automatic verification of that for the third column and, therefore,
we have found that several combination of matrix elements exist that qualify
$\bf O$ as the modulation matrix of an optimum polarimetric scheme, as we aimed
at demonstrating.

Real polarimeters, however, have some optics in between the modulator and the
analyzer. Very importantly, modern magnetographs like CRISP, VIP, IMaX, or
SO/PHI have one or several Fabry-P\'erot etalons. Such etalons can modify the
Mueller matrix that leads to a modulation matrix like that in Eq.\
(\ref{eq:modulmat}) and, hence, we must check whether or not the resulting
modulation matrix, ${\bf O}'$, remains optimum. To do that, let us model the
most general behavior of an etalon as a retarder ${\bf M}_3 = {\bf R}
(\theta_{\rm etalon}, \delta_{\rm etalon})$. Then, the final Mueller matrix of
the system is now ${\bf F} = {\bf M}_4 {\bf M}_3 {\bf M}_2 {\bf M}_1$ and its
first row (again disregarding the gain factor) is given by $F_{11} =1$,  
\begin{equation} \begin{array}{lll}
\label{eq:matelem}
F_{12} & = & M_{3,22} \cos\tau_i + M_{3,24} \sin \tau_i, \\
F_{13} & = & M_{3,22} \sin\rho_i  \, \sin\tau_i + M_{3,23} \cos\rho_i - M_{3,24} \sin\rho_i \, \cos\tau_i, \\
F_{14} & = & -M_{3,22} \cos\rho_i  \, \sin\tau_i + M_{3,23} \sin\rho_i + M_{3,24} \cos\rho_i \, \cos\tau_i.
\end{array}\end{equation}

We do not need any more matrix elements of ${\bf F}$ because the rows of the
new modulation matrix are $O'_{ij} = F_{1j} (\tau_i, \rho_i)$. Now, we only
need to find out four different combinations of the first and second
retardances that are solutions for Eqs.\ (\ref{eq:matelem}) with $|F_{1k}| =
1/\sqrt{3}$, where $k=2,3,4$. Equations (\ref{eq:matelem}) are transcendental
and, thus, have to be solved numerically. However, before proceeding with the
numerical exercise we can realize several features in the solutions. First,
the trivial cases, where $\delta_{\rm etalon} = 0$ (that is, no etalon exists
or it is not birefringent) or $\theta_{\rm etalon} = 0, \pi/2$, the orthogonal
directions of the analyzer axis, are indeed trivial because the effect of ${\bf
M}_3$ disappears and $O'_{ij} = O_{ij}$. Second, a number of periodicities can
be deduced from the equations structure: 
\begin{itemize}
\item If $\tau_{0}$ is a solution for the first of Eqs.\ (\ref{eq:matelem})
with $F_{12} = 1/\sqrt{3}$, then $\tau_{0} + (2k+1)\pi$, with $k$ integer, are
solutions for that equation when $F_{12} = -1/\sqrt{3}$ and vice versa.
\item If $\rho_{0}$ is a solution for the second or the third of Eqs.\
(\ref{eq:matelem}) with $F_{12} = 1/\sqrt{3}$, then $\rho_{0} + (2k+1)\pi$,
with $k$ integer, are solutions for that equation when $F_{12} = -1/\sqrt{3}$
and vice versa.
\item If $\rho_{0}$ is a solution for the second of Eqs.\ (\ref{eq:matelem})
with $F_{12} = 1/\sqrt{3}$, then $\rho_{0} + (2k+1)\pi/2$, with $k$ even
integer, are solutions for the third equation when $F_{12} = -1/\sqrt{3}$. When
$k$ is odd, then the solution for the third equation is when $F_{12} =
1/\sqrt{3}$ as well.
\end{itemize}

To solve the first of Eqs.\ (\ref{eq:matelem}) let us consider the function
\begin{equation}
\label{eq:ffunction}
f(\tau) = F_{12} - M_{3,22} \cos\tau - M_{3,24} \sin\tau,
\end{equation}
that has extrema where its derivative becomes zero. This occurs at $\tau_0$
where either $M_{3,24} = \sin\tau_0$ and $M_{3,22} = \cos\tau_0$ or $M_{3,24} =
-\sin\tau_0$ and $M_{3,22} = -\cos\tau_0$. These values imply that $f(\tau_{\rm
max}) = F_{12} + 1$ and $f(\tau_{\rm min}) = F_{12} - 1$. That is, the maximum
of the function is positive and the minimum is negative when $|F_{12}| =
1/\sqrt{3}$ (which is required for reaching optimum
efficiencies).\footnote{Note that we have selected one out of the infinite
solutions for the derivative of $f$ to be zero, but this is coherent with our
neglecting multiplicative, gain factors in the definition of Mueller matrices.}
Therefore, since $f$ is continuous, Bolzano's theorem ensures that a solution
exists in $(\tau_{\rm min}, \tau_{\rm max})$ and this enables us to find that
solution, for instance, through the bisector method. This has to be done just
once per value of $F_{12}$; the other value derives from the first of the above
specified properties.

As a summary, the first of Eqs. (\ref{eq:matelem}) has four solutions in
$[0,2\pi]$, each two belonging to one of the signs of $F_{12}$. For each of
these four retardances, $\tau_i$, four possibilities are open according to the
values of $F_{13}$ and $F_{14}$. These four solutions for $\rho_i$ in the
second and third of Eqs.\ (\ref{eq:matelem}) can be shown to be enclosed in the
following single expression, 
\begin{equation}
\label{eq:rhosol}
\cos\rho = \pm \frac{(n\mp m)}{\sqrt{3} \, (m^2+n^2)},
\end{equation}
where $n = M_{3,23}$ and $m = M_{3,22} \sin\tau - M_{3,24} \cos\tau$. A further
property of the solutions thus derives from Eq. (\ref{eq:rhosol}): if $\rho_0$
is a solution for the second and the third of Eqs.\ (\ref{eq:matelem}), then
$2\pi-\rho_0$, $\pi-\rho_0$, and $\pi + \rho_0$ are solutions as well.  

Therefore, the presence of an etalon modeled as a retarder is not a problem for
the two-LCVR-based polarimeter to be optimum. No matter the possible retardance
or orientation the etalon may have, we are always able to find out more than
four combinations of $\rho$ and $\tau$ that ensure theoretical polarimetric
efficiencies for all three Stokes parameters all equal to $1/\sqrt{3}$. In
practice, these new solutions can be achieved by simply tuning the acting
voltages of the two LCVRs.

\begin{figure}[!t]
\centering
\resizebox{\hsize}{!}{\includegraphics{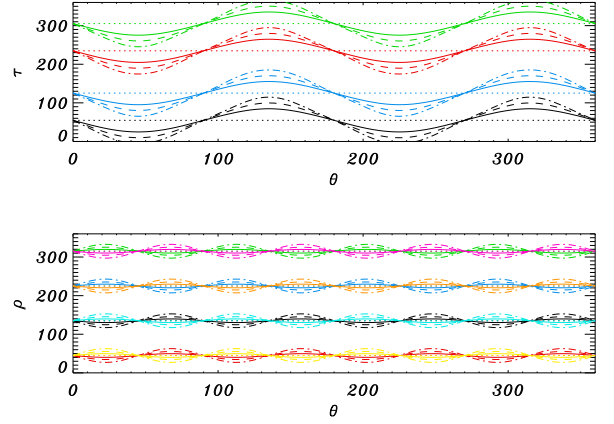}}
\caption{Retardances of the first (bottom) and second (top) LCVRs as functions
of the angle $\theta$ of the etalon orientation with respect to the $S_{2}$
positive direction. Only values between $0$ and $2\pi$ are displayed. Different
line types refer to different values of the etalon retardance (see text).}
\label{fig:retardances}
\end{figure}

Figure \ref{fig:retardances} displays the LCVR retardances in $[0,2\pi]$ that
guarantee optimum performance as functions of the orientation angle of the
etalon. Values for the second LCVR are in the top panel and those for the first
one are in the bottom panel.\footnote{Note that retardances larger than $2\pi$
may be needed for design convenience. Their values are easily deducible from
the above-mentioned properties.} Different colors correspond to different
solutions; different line types correspond to different values of the assumed
etalon retardance: 0\degree (dotted), 30\degree (solid), 45\degree (dashed),
and 60\degree (dashed-dotted). As commented on before, when $\theta_{\rm
etalon} = 90\degree$, both $\rho$ and $\tau$ recover the same value as if the
etalon were absent. Moreover, both retardances are periodic with $\theta$, with
periods $\pi/4$ ($\rho$) and $\pi/2$ ($\tau$). It is also interesting to note
that four out of the eight solutions for $\rho$ are equal to the other four but
phase shifted by $\pi/4$.

Now, it is a little tedious but easy to demonstrate that, regardless of how
many, mirrors can be introduced in the optical path between the modulator and
the analyzer (as in real instruments) without affecting the maximum
polarimetric efficiencies, provided they all are perpendicular to the optical
axis plane. From the Mueller matrix of a single mirror, one can realize
\citep{collett1992} that the matrix of such a mirror train, no matter the
angles between them, keeps always the shape
\begin{equation}
\label{eq:mirrortrain}
{\bf E} = \left( \begin{array}{cccc}
a & b & 0 & 0 \\
b & a & 0 & 0 \\
0 & 0 & c & d \\
0 & 0 & -d & e \end{array} \right).
\end{equation}
The elements of the first row in the new final Mueller matrix of the system
become $F'_{1j} = (a+b) F_{1j}$, $j=1,2,3,4$. That is, the final modulation
matrix remains the same as before introducing the mirrors but scaled by a gain
factor that can be disregarded as we have been doing for all the treatment.
Therefore, we can conclude that optimum efficiencies can still be achieved with
as many mirrors as needed. Since a mirror is indeed a combination of a retarder
and a partial polarizer \citep[the $d$ element is zero for the latter; see
e.g.][]{1994KAP...book...S}, the same conclusion can be reached for whatever
differential absorption effects for the orthogonal polarization states that may
be located between the modulator and the analyzer. Therefore, if, for example,
the etalon or the diffraction grating of the instrument display different
transmittances for orthogonally polarized beams the polarimetric efficiencies
can still attain maximum values.

\section{Instrument-induced inaccuracies}
\label{sec:instabilities}

Now that we know that our magnetograph can reach optimum polarimetric
performance, let us study the behavior of this particular instrument against
instabilities in its main optical elements. Photon noise is not the only harm
for magnetographic or tachographic measurements. Instabilities of different
types like those in the temperature or in the tuning voltage of both the LCVRs
and the etalon, or roughness in their final thicknesses, can induce
inaccuracies. For single measurements,
the inaccuracies can imply errors in magnetic field or absolute wavelength
calibration. For time series like those needed in helioseismological studies,
such inaccuracies may avoid detection of some particular oscillatory modes. An
assessment on such inaccuracies is therefore in order for clearly defining
design tolerances of these instrumental quantities. Such tolerances should
ensure the fulfillment of the scientific requirements of the instrument. 

\subsection{Polarimetric inaccuracies}
\label{sec:tempinac}

An important example of a scientific requirement is the polarimetric accuracy
of the system. By such we understand any of the inverse signal-to-noise ratios
for $S_2$, $S_3$, or $S_4$ as defined in Eq.\ (\ref{eq:snratio}). Following our
general assumption that these $(S/N)_i$ are intended to be the same by design,
aiming at a $S/N$ of 1700 is equivalent to require the system to have a
polarimetric accuracy of $10^{-3}$. Temperature, voltage, and other instabilities
and defects of the LCVRs lead to changes in the retardances that, on their
turn, induce modulation and demodulation changes. Such changes can be seen as
cross-talk between the Stokes parameters that drive to covariances in the
magnetographic measurements (\citeauthor{2008ApOpt..47.2541A}
\citeyear{2008ApOpt..47.2541A}; see also an interesting discussion on
seeing-induced cross-talk in \citeauthor{2011arXiv1107.0367C}
\citeyear{2011arXiv1107.0367C}). Since our modulation matrix
(\ref{eq:modulmat}) is analytical, we attempt an
analytical approach to the study of the
effect of these retardance changes onto the polarimetric accuracy of the
system.   

Let ${\bf D}$ be the demodulation matrix of the instrument that always exists.
Thus, ${\bf OD} = {\bf DO} = \bbbone$. The measured Stokes vector is then given
by ${\bf S} = {\bf D} {\bf I}_{\rm meas}$, where ${\bf I}_{\rm meas}$ stands
for the four intensity measurement vector of each modulation cycle. If we
linearly perturb matrix ${\bf D}$ as a consequence of a small perturbation in
the LCVR retardances ($\sigma_{\rho_{k}}, \sigma_{\tau_{k}}, \, k= 1, 2, 3,
4$), then the Stokes vector becomes ${\bf S} + {\bf S}^{\prime} = ({\bf D} +
{\bf D}^{\prime)} {\bf I}_{\rm meas}$, where the perturbed demodulation matrix
${\bf D}^{\prime}$ is given by
\begin{equation}
\label{eq:dprime}
{\bf D}^{\prime} = \sum_{k=1}^4 \left( \frac{\partial {\bf D}}{\partial \rho_{k}} \sigma_{\rho_{k}} + 
\frac{\partial {\bf D}}{\partial \tau_{k}} \sigma_{\tau_{k}} \right).
\end{equation}
It is obvious that the polarimetric accuracy requirement directly implies that
none of the four elements of ${\bf S}^{\prime} = {\bf D}^{\prime} {\bf I}_{\rm
meas}$ can be greater than $10^{-3}$. To fulfill that
requirement, let us then study how the perturbation in
the retardances can be produced and which are the tolerances for the instrument
quantities whose fluctuations produce them.

If we call $\delta_{L}$ anyone of the retardances, we have by definition that
\begin{equation}
\label{eq:deltadef}
\delta_{L} = \frac{\beta t}{\lambda_0},
\end{equation}
where $\beta = n_{\rm e} - n_{\rm o}$ is the birefringence, i.e., the
difference between the extraordinary and the ordinary refractive indices of the
liquid crystal, and $t$ stands for its geometrical thickness. Therefore, it is
evident that 
\begin{equation}
\label{eq:deltaerror1}
\frac{\sigma^2_{\delta_L}}{\delta^2_L} = \frac{\sigma^2_{\beta}}{\beta^2} + \frac{\sigma^2_{t}}{t^2},
\end{equation}
that is, the relative inaccuracy in the LCVR retardance is the square root of
the sum of the square relative inaccuracies in the birefringence and in the
geometrical thickness at the operating wavelength. Let us ascribe for
convenience any possible local fabrication defect in the LC like an air bubble
to the thickness inaccuracy, so that birefringence can be considered spatially
constant for the whole device. 

\begin{figure}[!t]
\centering
\resizebox{\hsize}{!}{\includegraphics{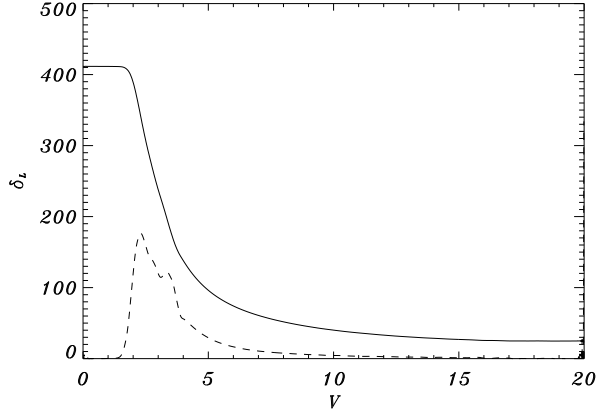}}
\caption{Retardance (in degree) of a specific nematic LCVR as a function of the
acting voltage (in volts; solid line) and its derivative with inverted sign
(dashed line).}
\label{fig:retvolt}
\end{figure}

Variations in the birefringence can be produced by either variations in the
LCVR temperature, the acting voltage, or both. Hence, one can write
\begin{equation}
\label{eq:bireferror}
\sigma^2_{\beta} = q_T^2 \sigma_T^2 + q_V^2 \sigma_V^2,
\end{equation}
where $q_T$ and $q_V$ are values of the (partial) derivatives of $\beta$ with
respect to $T$ and $V$ at the given values of voltage and temperature,
respectively. Since we indeed have calibrations of the $\delta_{L}$ dependences
rather than those of $\beta$, we better rewrite Eq.\ (\ref{eq:deltaerror1}) as
\begin{equation}
\label{eq:deltaerror2}
\sigma^2_{\delta_L} = m_T^2 \, \sigma^2_{T} + m_V^2 \, \sigma^2_{V} + \frac{\delta_L^2}{t^2} \, \sigma^2_{t},
\end{equation}
where $m_T^2$ and $m_V^2$ have a clear meaning and can be deduced from
calibrations. According to \citet{2011SoPh...268...57M}, based on data by
\citet{2007ApOpt..46..689H}, 
\begin{equation}
\label{eq:emet}
m_T = -1.16 + 0.305 V - 0.02 V^2,
\end{equation}
for $V < 8$ volt and $0$ otherwise, and $m_V$ can be obtained from data like
those displayed in Fig.\ \ref{fig:retvolt} where $\delta_L$ (solid line) and
its derivative (dashed line; inverted sign) are plotted as functions of the
acting voltage for a particular LCVR.

To get a numerical estimation of the value of real tolerances, that is, of the
maximum $\sigma_T$, $\sigma_V$, and $\sigma_t$ affordable
in real instruments in order not to have the ${\bf S}^{\prime}$-elements
greater than $10^{-3}$, we shall use IMaX parameters\footnote{The combination
for IMaX retardances was $\rho = [315, 315, 225, 225]$ and $\tau = [305.264,
54.736, 125.264, 234.736]$, in degrees. Their corresponding voltages were
$V(\rho) = [2.535, 2.535, 3.112, 3.112]$ and $V(\tau) = [2.4, 9.0, 4.3, 2.9]$,
in volts.} plus the analytic expressions for ${\bf D}$ and its partial derivatives.\footnote{An IDL program that evaluates such analytic expressions is available upon request. Direct (by hand) evaluation is so tedious that only with the help of software applications like Mathematica such expressions can be obtained.} The last term in Eq.\ (\ref{eq:deltaerror2}) may
vary spatially and is, thus, responsible for the pixel-to-pixel variations 
of the retardance but can easily be calibrated if needed. Indeed, roughness in the
device thickness is a fabrication specification and can be checked upon
delivery from the manufacturer. Thickness inaccuracies may produce locally
significant effects \citep[e.g.,][]{2010EPJWC...505002A}. 
Our own estimation, using Eqs.\ (\ref{eq:dprime}, \ref{eq:deltaerror2}) indicate that relative errors in the thickness larger than 6 \% induce perturbations of the Stokes vector that are larger than
the polarimetric accuracy. Therefore, should this be the only instrumental
instability, specific pixel-to-pixel calibration of ${\bf D}$ would be needed
if the relative roughness is larger than 6~\%. Note that, since the typical
thicknesses of LCVRs are of the order of micrometers, a roughness less than 6\%
may mean a stringent requirement for the manufacturer of the order of tens of
nanometers. 

Using again Eqs.\ (\ref{eq:dprime}, \ref{eq:deltaerror2}), we find out that instabilities larger than $\sigma_T
= 600$ mK or $\sigma_V = 1.5$ mV deteriorate the polarimetric accuracy below
the required $10^{-3}$. 
These two tolerances and that for the roughness have been calculated after
assuming that each instability is acting individually. According to Eq.\
(\ref{eq:deltaerror2}), the final uncertainty in the retardance stems from 
the three sources simultaneously. Hence, a safety
reduction factor of $\sqrt{3}$ (assuming all the three contribute the same) is
advisable. Therefore, in the end, the final tolerance specification for our instrument to
reach a polarimetric accuracy of $10^{-3}$ is 300 mK for temperature, 1 mV for
voltage, and a 4 \% for LCVR roughness.

\subsection{Magnetographic inaccuracies}
\label{sec:magnetinac}

The retardance perturbations of Eq.\ (\ref{eq:deltaerror2}) induce changes in the
maximum polarimetric efficiencies of the instrument. Such changes necessarily imply
modifications in the magnetographic measurements. The modifications might jeopardize the quality of the results. Imagine, for instance, that a requirement on the repeatability of $B_{\rm lon}$ and $B_{\rm tran}$ applies because we are interested on a measurement time series: a calculation of tolerances in the instrument parameters ($T$, $V$, roughness, etc.) that ensure the fulfillment of the magnetographic repeatability is in order. 

Since no explicit dependence of $B_{\rm lon}$ and $B_{\rm tran}$ on $\varepsilon_{i}$ exists, we cannot analytically gauge the induced inaccuracies in any magnetographic measurement. Nevertheless, we can obtain a hint on the global by studying the specific variations of $\delta (B_{\rm lon})$ and $\delta (B_{\rm tran})$, the minimum detectable values of such magnetograph quantities.

Error propagation in Eqs.\ (\ref{eq:uncerblon}) and (\ref{eq:uncertran}) readily gives
\begin{equation}
\label{eq:erruncerblon}
\sigma^2_{{\delta(B_{\rm lon})}} = \frac{{\delta^2(B_{\rm lon})}}{4 \varepsilon_4^2} \, \sigma^2_{\varepsilon^2_{4}}
\end{equation}
and 
\begin{equation}
\label{eq:erruncertran}
\sigma^2_{{\delta(B_{\rm tran})}} = \frac{{\delta^2(B_{\rm tran})}}{16 \varepsilon_2^2} \, \sigma^2_{\varepsilon^2_{2}} = \frac{{\delta^2(B_{\rm tran})}}{16 \varepsilon_3^2} \, \sigma^2_{\varepsilon^2_{3}}.
\end{equation}
According to \cite{2000ApOpt..39.1637D}, the maximum polarimetric efficiencies that can be reached by any system are
\begin{equation}
\label{eq:defeff}
\varepsilon^2_{{\rm max},i} = \frac{\sum_{j=1}^4 O^2_{ji}}{N_p},
\end{equation}
where $O_{ji}$ are the matrix elements of ${\bf O}^{\rm T}$, the transpose of $\bf O$. For a system with a modulation matrix like that in Eq.\ (\ref{eq:modulmat}), it is easy to see that the efficiency inaccuracies ensuing the thermal instabilities are
\begin{equation}
\label{eq:errepsilon1}
\sigma^2_{\varepsilon^2_{{\rm max},1}} = 0,
\end{equation}
as a consequence of using normalized Mueller matrices, and
\begin{equation}
\label{eq:errepsilon2}
\sigma^2_{\varepsilon^2_{{\rm max},2}} = \frac{1}{16} \sum_{j=1}^4 \sin^2 2\tau_j \,\, \sigma^2_{\tau_j},
\end{equation}
\begin{eqnarray}
\label{eq:errepsilon3} 
\sigma^2_{\varepsilon^2_{{\rm max},3}} & = & \frac{1}{16} \left\{  \sum_{j=1}^4 \sin^2 2\rho_j \,\, \sin^4 \tau_j \,\, \sigma^2_{\rho_j} \right. + \nonumber \\ 
& & \left. \sum_{j=1}^4 \sin^4 \rho_j \,\, \sin^2 2\tau_j \,\, \sigma^2_{\tau_j} \right\}, 
\end{eqnarray}
\begin{eqnarray}
\label{eq:errepsilon4} 
\sigma^2_{\varepsilon^2_{{\rm max},4}} & = & \frac{1}{16} \left\{  \sum_{j=1}^4 \sin^2 2\rho_j \,\, \sin^4 \tau_j \,\, \sigma^2_{\rho_j} \right. + \nonumber \\
& & \left. \sum_{j=1}^4 \cos^4 \rho_j \,\, \sin^2 2\tau_j \,\, \sigma^2_{\tau_j} \right\}, 
\end{eqnarray}
where $\sigma^2_{\tau_j}$ and $\sigma^2_{\rho_j}$ are the variances of the LCVR retardances and where we have neglected the influence of possible time-exposure differences or instabilities like those caused by a rolling-shutter detector or a non-ideal repeatability of a mechanical shutter. Such influences can be estimated separately for the specific instrument and directly scale the effective exposure time (either per pixel or per frame).

Using the values for IMaX, and assuming maximum efficiencies, instabilities of 0.3 K or of 1.1 mV produce a 5 \% repeatability error in the threshold for $B_{\rm lon}$. A comparison of Eqs.\ (\ref{eq:erruncerblon}) and (\ref{eq:erruncertran}) readily tells that the effect is 2 times smaller on the relative repeatability error in the threshold for $B_{\rm tran}$ but since the threshold itself is 16 times larger, the absolute error is in the end 8 times larger as well.

\subsection{Velocity inaccuracies}
\label{sec:velinacc}

Error propagation in Eq.\ (\ref{eq:tachometer}) yields
\begin{equation}
\label{eq:errvel1}
\sigma_{v_{\rm LOS}}^2 = \left( \frac{\partial v_{\rm LOS}}{\partial \delta\lambda} \right)^2 \sigma_{\delta\lambda}^2 + \left( \frac{\partial v_{\rm LOS}}{\partial \lambda_0} \right)^2 \sigma_{\lambda_0}^2 + \sum_k\left( \frac{\partial v_{\rm LOS}}{\partial S_{1,k}} \right)^2 \sigma_{{1,k}}^2
\end{equation}
with $k = -9, -3, +3, +9$. Uncertainties in the spectral resolution come from uncertainties in the etalon spacing and related fabrication details that produce etalon roughness. Errors in the central wavelength come from the etalon tuning that mostly depends on the ambient temperature, $T$, and on the tuning voltage, $V$:
\begin{equation}
\label{eq:errcenlam}
\sigma_{\lambda_0}^2 = k_T^2 \sigma_T^2 + k_V^2 \sigma_V^2,
\end{equation}
where $k_T$ and $k_V$ are constants that give the (linear) dependence of $\lambda_0$ on $T$ and $V$. Finally, uncertainties in the Stokes $S_1$ samples come both from pure photon noise, as in
any photometric measurement, and from etalon tuning uncertainties. Since the (inexplicit) dependence of $S_{1,k}$ on $\lambda_0$ is non linear, let us linearize it (that is, introduce a small perturbation and take the first approximation) and write
\begin{equation}
\label{eq:sigmak}
\sigma_{1,k}^2 = \sigma_1^2 + s_{1,k}^2 \sigma_{\lambda_0}^2,
\end{equation}
where we have assumed that the photometric contribution is equal for all the samples
and indeed equal to the photon noise as calculated in the continuum;  $s_{1,k}$ stand for the
derivatives of the Stokes $S_1$ profile at the corresponding wavelengths.

After a tedious but straightforward algebra, Eq.\ (\ref{eq:errvel1}) can be recast as
$$\sigma_{v_{\rm LOS}}^2 = \frac{v_{\rm LOS}^2}{(\delta\lambda)^2} \sigma_{\delta\lambda}^2 + 
\frac{4c^2(\delta\lambda)^2}{\pi^2 \lambda_0^2} \frac{2\sigma_1^2}{\Delta} +$$
\begin{equation}
\label{eq:errvel2}
\left[ \frac{v_{\rm LOS}^2}{\lambda_0^2} + \frac{4c^2(\delta\lambda)^2}{\pi^2 \lambda_0^2} \frac{d_1 + d_2}{\Delta^2} \right] (k_T^2 \sigma_T^2 + k_V^2 \sigma_V^2),
\end{equation}
where $\Delta$ is defined in Sect.\ \ref{sec:uncer}, $d_1 = (S_{1,+9} - S_{1,-3})^2 (s_{1,-9}^2 + s_{1,+3}^2)$ and $d_2 =  (S_{1,-9} - S_{1,+3})^2 (s_{1,+9}^2 + s_{1,-3}^2)$. Hence, clear contributions to the final LOS velocity uncertainty can be discerned from the etalon roughness, the photon noise, the etalon temperature instability, and the etalon voltage instability.

Quantitative estimates of the various terms in Eq.\ (\ref{eq:errvel2}) can be
made by using the FTS spectrum by \citet{1987ftp...book...B} to evaluate
$\Delta$, $d_1$, and $d_2$ for a given spectral line and a given instrument.
Let assume the HMI and SO/PHI Fe~{\sc i} line at $\lambda_0 = 617.3$ nm and a
spectral resolution of the etalon of $\delta \lambda = 10$ pm. The first term
has a clear impact on the tachographic results: the etalon relative roughness
is directly translated into the same $v_{\rm LOS}$ relative uncertainty. In
other words, we cannot expect better accuracy in the line-of-sight velocity
(when measured with the Fourier tachometer formula) than that limited by the
etalon relative roughness; this means that a mere 0.1 pm, rms resolution
uncertainty induces 10 m$\,$s$^{-1}$ errors for speeds of 1 km$\,$s$^{-1}$. The
second term coincides with the right-hand side of Eq.\
(\ref{eq:noiseinducederrvel}) and has been discussed already in Sect.\
\ref{sec:uncer}. Since the ratio between the second and the first terms within
brackets is of the order of 4$\cdot$10$^9$ for velocities of up to 5
km$\,$s$^{-1}$, it is the second one what really matters in the estimation;
this means that the dependence of the profile shape on the central wavelength
of the line is really important. If we use the same IMaX values of $k_T = 2.52$
pm/K and $k_V = 3.35\cdot$10$^{-2}$ pm/V for the SO/PHI etalon, Eq.\ (\ref{eq:errvel2})  gives 
instabilities of the order of 3.3 mK or 0.25 V induce the same LOS velocity
uncertainty of 4 km$\,$s$^{-1}$ quoted above for pure photon noise. Another way of seeing the same effect can be explained by saying that a 100
m$\,$s$^{-1}$ uncertainty is produced by either a 45 mK or a 3.4 V
instabilities. These uncertainties are really important when stability during
given periods of time of the instrument is required as for helioseismic
measurements. For single shots, uncertainties in temperature or voltage imply
thresholds for accurate absolute wavelength (velocity) calibration. The
importance of having included the measurement technique in this error budget
analysis is clear: should one have simply used the $k_{V}$ and $k_{T}$
calibration constants above and the Doppler formula an uncertainty of just 55
m$\,$s$^{-1}$ would have been obtained. Hence, the uncertainty would have been
underestimated by a factor almost 2.

\section{Summary and conclusions}
\label{sec:conclu}

An assessment study on the salient features and properties of solar
magnetographs has been presented. An error budget procedure has been followed.
Special care has been devoted in including photon-induced and
instrument-induced noise as well as specific measurement technique
contributions to the final variances. We have first discussed the effect of
random noise in the measurements and deduced useful formulae --general for
every device-- that provide some minimum detectable parameters like the degree
of polarization of light, the longitudinal and transverse components of the
magnetic field, and the line-of-sight velocity. The detection thresholds are
given as functions of the polarimetric efficiencies of the instrument and of
the signal-to-noise ratio of the observations. (As a proposal, we have
suggested as well to use the $S/N$ for the Stokes intensity as {\em the}
signal-to-noise ratio for the instrument.) When the random noise is
photon-induced, we have calculated as well the relative uncertainty in the
magnetographic and tachographic quantities. Secondly, an analysis is presented
for those instruments based on two nematic liquid crystal variable retarders as
a polarization modulator and a Fabry-P\'erot etalon as the spectrum analyzer.
Although specific for these magnetographs, the methodology can easily be
followed by others in order to characterize their capabilities and accuracies.
We have demonstrated that this type of instrument can indeed reach theoretical
maximum polarimetric efficiencies because solutions always exist for the
retardances of the two LCVRs that ensure such efficiencies, hence optimizing
the detection thresholds and the relative uncertainties. Very remarkably, the
existence of such solutions is independent of the optics that is in between the
polarization modulator and the analyzer. Neither retarders nor partial
polarizers or mirrors (the most commonly used devices) alter that property. The
LCVR {\em optimum} retardances do depend in such pass-through optics but can be
fine-tuned according to the polarizing properties of the optics. A number of
rules and periodicity properties of the required retardances have also been
deduced. These polarimeters have modulation and demodulation matrices that are
explicitly calculated through an IDL procedure that is available upon request. Thirdly, error propagation has
yielded equations relating the variances of the measured Stokes vector and the
solar physical quantities and instrument parameters, hence providing a bridge
between scientific requirements and instrument design specifications. The
analytic character of this particular type of instrument has also allowed the
quantitative estimation of the mentioned uncertainties. Hopefully, the
discussion presented in this paper excites (and helps) further diagnostics of
other instruments.

\acknowledgments

This work has been partially funded by the Spanish Ministerio de Ciencia e Innovaci\'on, 
through Projects No. AYA2009-14105-C06 and AYA2011-29833-C06, and Junta de Andaluc\'{\i}a, through Project 
P07-TEP-2687, including a percentage from European FEDER funds.

\bibliography{/Users/jti/Dropbox/BIBTEX/Biblioteca_de_citas}

\end{document}